\documentclass[aps,preprint,prd,nofootinbib,epsfig]{revtex4-1}
\usepackage[latin1]{inputenc}
\usepackage{amssymb}
\usepackage{amsmath}
\usepackage{graphicx}
\usepackage{fancyhdr}
\newcommand{\be}{\begin{equation}}
\newcommand{\ee}{\end{equation}}
\newcommand{\bea}{\begin{eqnarray}}
\newcommand{\eea}{\end{eqnarray}}
\newcommand{\bes}{\begin{subequations}}
\newcommand{\ees}{\end{subequations}}

\newcommand{\bc}{\begin{center}}
\newcommand{\ec}{\end{center}}

\begin{document}

\title{The Higgs sector of the SUSY reduced 3-3-1 model }

\author{J. G. Ferreira Jr.,  C. A. de S. Pires, P. S. Rodrigues da Silva, A. Sampieri}
\email{josegeilson@fisica.ufpb.br \\  cpires@fisica.ufpb.br \\ psilva@fisica.ufpb.br \\asampieri@fisica.ufpb.br}
\affiliation{{ Departamento de
F\'{\i}sica, Universidade Federal da Para\'\i ba, Caixa Postal 5008, 58051-970,
Jo\~ao Pessoa, PB, Brasil}}

\date{\today}

\begin{abstract}
A supersymmetric version of the recently proposed reduced minimal 3-3-1 model is considered and its Higgs sector is investigated. We focus on  the mass spectrum of the lightest scalars of the model. We show that Higgs mass of 125 GeV requires substantial radiative corrections. However,   stops may develop small mixing and  must have mass around TeV. Moreover, some soft SUSY  breaking terms may lie  at the electroweak scale, which alleviates some tension  concerning fine tuning of the related parameters. The lightest  doubly charged scalar may have  mass around few hundreds of GeV, which can be probed at the LHC, while the remaining scalars of the model have masses at TeV scale. 
\\ 
PACS: 12.60.Cn; 12.60.Jv; 14.80.Da.
\end{abstract}

\maketitle

\section{Introduction}
It seems that the ultimate block of the standard model (SM) of particle physics was finally detected in CERN  Large Hadron Collider (LHC) by ATLAS and CMS experiments\cite{LHC}. Both experiments  recently reported the detection of  a new particle with mass around $125$GeV.  Fitting to all available collider data  suggest that the discovered particle is, indeed, the elusive  Higgs boson.

 It is well accepted that the standard model (SM) is not the final answer in particle physics. A couple of experimental results  as, for example, neutrino oscillation~\cite{neutrinos},  dark matter~\cite{darkmatter}, etc., require extension of the SM. From the theoretical side, the SM also suffers from incompleteness once it is  not able to explain problems like  the hierarchy problem,  family replication, and so on. Thus we see that at the moment we have experimental and theoretical reasons to go beyond the SM.

Among the alternatives to SM, there is Supersymmetry (SUSY), that has been fascinating physicists for more than three decades. The reasons behind the great interest in SUSY  lies in the fact that it is an attractive solution to the hierarchy problem by entangling fermions and bosons, and then justifying the presence of fundamental scalars in the spectrum. Besides, it allows  unification of the coupling constants and predicts a light Higgs boson to engender the electroweak symmetry breaking~\cite{susyreviews}.  In the most popular low energy SUSY model, the minimal supersymmetric standard model (MSSM),  a 125~GeV Higgs mass requires substantial loop contributions once a SM-like Higgs mass is upper bounded by $ M_Z \cos 2\beta$ at tree-level.  A great number of papers analyzing MSSM's parameter space to reproduce LHC data was published in the last year, showing that, although very restrictive, there is still enough room in the parameter space for a light Higgs boson~\cite{susypapers}.

On the other hand, in a  class of gauge models based on the $SU(3)_C \times SU(3)_L \times U(1)_N$ (331) gauge symmetry,   anomaly cancellation requires the existence of at least three families of fermions~\cite{ffv}. Consequently, the supersymmetric versions of these gauge models would  solve the hierarchy problem and  family replication altogether.  Recently it was shown that the so-called minimal  331 models may be implemented with two Higgs triplets only~\cite{reduced}. The model is very short and predictive in its scalar spectrum, compared to the original version~\cite{ffv}. It was called reduced 331 model and some of its phenomenology was developed in Ref.~\cite{pheno}. Curiously, the scalar content that survives the reduction has the appropriate quantum numbers to be supersymmetrized, without the need of extra multiplets employed in the first SUSY versions of the minimal 331 model~\cite{DuongMa,marcao}. Actually, while we were working on this model, the authors in Ref.~\cite{susylong} also observed this, although their model, as well as results and conclusions, differ in some crucial points from ours as we will remark later. There is an additional issue tha motivates the supersymmetrization of this model, namely, the non-supersymmetric version of minimal 331 has no natural candidate for cold dark matter~\footnote{Although there is a claim in Ref.~\cite{mauro} that a stable self-interacting dark matter candidate exists in the minimal 331 model, it is easy to show that the lack of a symmetry to guarantee its stability, allows us to write down effective operators that are not sufficiently supressed to avoiding its decay.}, while such a candidate is natural in its SUSY version once R-parity is present.
 
In this work we develop the SUSY version of the reduced 331 model. We focus on the scalar spectrum of the model and give particular emphasis on the Higgs boson that arises in its spectrum. Our main result is that the stability of the vacuum imposes a Higgs mass upper bounded $< 90$~GeV at tree level. We calculated the radiative corrections that lift this mass to  $125$~GeV. We also obtained the masses of the lightest charged scalars.

\section{The essence of the reduced SUSY331 Model}
In order to implement the supersymmetric version of a certain model, we have  to promote its fields to superfields. In this way, the leptons in the  reduced SUSY 331 (RSUSY331) model  compose three  chiral left-handed lepton superfields  denoted by, 
\begin{equation}
\begin{array}{cc}
\hat{L}_L=\left(\begin{array}{c}
\hat{\nu}_l \\  \hat{l}\\ \hat{l}^c
\end{array}\right)_L & \sim({\bf1},{\bf3},0),
\end{array}
\label{leptonsuperfields}
\end{equation}
where $l=e,\mu ,\tau$. Notice that the presence of a right-handed component of the charged lepton field in the multiplet, allows us to dispose of singlet right-handed leptons.

For the chiral left-handed quark superfields,  the first family comes in triplet representation, while the second and third families come in anti-triplet representation, while their right-handed partners are arranged in singlets as denoted below,
\begin{eqnarray}
&& \begin{array}{cc}
\hat{Q}_{1L}=\left(\begin{array}{c}
 \hat{u}_1 \\ \hat{d}_1 \\ \hat{J}_1
\end{array}\right)_L & \sim({\bf3},{\bf3},\frac{2}{3})
\end{array}\,\,\,,\,\, \begin{array}{cc}
\hat{Q}_{iL}=\left(\begin{array}{c}
 \hat{d}_i \\ -\hat{u}_i \\ \hat{J}_i
\end{array}\right)_L & \sim({\bf3},{\bf3}^*,-\frac{1}{3}),
\end{array}\nonumber \\
&&
 \begin{array}{ccc}
\hat{u}^c_{1L} \sim\;({\bf3},{\bf1},-\frac{2}{3}), & \hat{d}^c_{1L}
\sim\;({\bf3},{\bf1},+\frac{1}{3}), & \hat{J}^c_{1L}
\sim\;({\bf3},{\bf1}-\frac{5}{3}),
\end{array}\nonumber \\
&& \begin{array}{ccc}
\hat{u}^c_{iL} \sim\;({\bf3},{\bf1},-\frac{2}{3}), & \hat{d}^c_{iL}
\sim\;({\bf3},{\bf1},+\frac{1}{3}), & \hat{J}^c_{iL}
\sim\;({\bf3},{\bf1},+\frac{4}{3}),
\end{array}
\label{quarksfamilies}
\end{eqnarray}
with $i=2,3$. Here, the extra quarks, $J_i$ and $J_1$ are exotic ones with electric charges $+4/3$ and $-5/3$, respectively.

The scalar sector of reduced  331 model is composed by two Higgs triplets . Consequently, anomaly cancellation requires that its supersymmetric version  possess four chiral left-handed Higgs superfields,
\begin{eqnarray} 
&& \,\,\hat{\rho} = 
      \left( \begin{array}{c} \hat{\rho}^{+} \\ 
                  \hat{\rho}^{0} \\
                  \hat{\rho}^{++}          \end{array} \right)
\sim ({\bf1},{\bf3},+1),\quad 
\,\,\,\,\,\hat{\chi} = 
      \left( \begin{array}{c} \hat{\chi}^{-} \\ 
                  \hat{\chi}^{--} \\
                  \hat{\chi}^{0}          \end{array} \right)
\sim ({\bf1},{\bf3},-1),\nonumber \\
&& \hat{\rho}^{\prime} = 
      \left( \begin{array}{c} \hat{\rho}^{{}_{\prime}-} \\ 
                  \hat{\rho}^{{}_{\prime}0} \\
                  \hat{\rho}^{{}_{\prime}--}          \end{array} \right)
\sim ({\bf1},{\bf3}^{*},-1),\quad 
\hat{\chi}^{{}_{\prime}} = 
      \left( \begin{array}{c} \hat{\chi}^{{}_{\prime}+} \\ 
                  \hat{\chi}^{{}_{\prime} ++} \\
                  \hat{\chi}^{{}_{\prime} 0}          \end{array} \right)
\sim ({\bf1},{\bf3}^{*},+1).
\label{scalarsuperfields}
\end{eqnarray}

These scalar superfields are not enough to render the correct mass pattern for all fermion fields through the superpotential though. Nevertheless, we recall that this class of 331 models possesses a Landau pole around 4 to 5~TeV~\cite{landaupole}, becoming strongly interacting before that point. Throughout this work we assume that the highest energy scale where the model is found to be pertubatively reliable is $\Lambda=5$~TeV. This is a welcome information since it allows us to make use of effective operators to complement that part of  the mass spectrum not obtained from the renormalizable superpotential. That being said, the superpotential~\footnote{Our superpotential (and soft SUSY breaking terms) is very distinct from that in Ref.~\cite{susylong}, mainly because we assume the usual R-parity as the MSSM, since lepton and baryon number are conserved in all interactions due to the association of leptonic number to some fields, called bileptons. Such bileptons are those scalar and vector fields which connect the first or second component of lepton fields in the triplets to the third one.} capable of generating the correct masses of the charged fermions in the RSUSY331  model is composed by the following terms,
\begin{equation}
\hat{f}= \hat{f}^{\prime} +  \hat{f}_{E.O.},
\label{superpotencial}
\end{equation}
where,
\begin{eqnarray}
\hat{f}^{\prime} &=& \lambda^J_{11} \hat{Q}_{1L} \hat{\chi}^{{}_{\prime}} \hat{J}^c_{1L} + \lambda^J_{ij}\hat{Q}_{iL}\hat{\chi} \hat{J}^c_{jL} + \lambda^{{}_{\prime}}_{1a} \hat{Q}_{1L} \hat{\rho}^{{}_{\prime}} \hat{d}^c_{aL} 
\nonumber \\ & &\mbox{}
 + \lambda^{{}_{\prime}{}_{\prime}}_{ia}\hat{Q}_{iL}\hat{\rho} \hat{u}^c_{aL} + \mu_{\rho}\hat{\rho}\hat{\rho}^{{}_{\prime}} + \mu_{\chi}\hat{\chi}\hat{\chi}^{{}_{\prime}},
 \label{firstpart}
\end{eqnarray}
and,
\begin{eqnarray}
 &&\hat{f}_{{}_{E.O.}}= \dfrac{k_{u{}_{1a}}}{\Lambda} \varepsilon_{nmp}\left(\hat{Q}_{1Ln}\hat{\rho}_m \hat{\chi}_p\right)\hat{u}^c_{aL} \nonumber \\ & &
 +\dfrac{k_{d{}_{ia}}}{\Lambda}\varepsilon_{nmp}\left(\hat{Q}_{iLn}\hat{\rho}^{{}_{\prime}}_m \hat{\chi}^{{}_{\prime}}_p \right)\hat{d}^c_{aL}  
+ \dfrac{k_{l}}{\Lambda}\left(\hat{L}_L\hat{\rho}^{{}_{\prime}}\right)  \left(\hat{L}_L \hat{\chi}^{{}_{\prime}} \right).
\label{E.O}
\end{eqnarray}
As before, $i,\,j=2,\,3$ and $a = 1,\,2,\,3$ are family index labels. 
 
The soft SUSY breaking terms of the model are given by,
\begin{eqnarray} 
{\cal L}_{soft} &=& - \frac{1}{2} \left[m_{\lambda_c}\lambda^a_c\lambda^a_c
+ m_{ \lambda} \left( \lambda_{a} \lambda_{a} \right) 
+ m^{ \prime} \lambda \lambda +h.c.\right] \nonumber \\ && - m_{L}^{2} \tilde{L}^{ \dagger} 
\tilde{L} - m_{Q_1}^{2} \tilde{Q}^{ \dagger}_{1} \tilde{Q}_{1} +
m_{u_{\alpha}}^2 \tilde{u}^{\dagger}_{\alpha} \tilde{u}_{\alpha} - m_{d_{\alpha}}^2 \tilde{d}^{\dagger}_{\alpha} \tilde{d}_{\alpha} \nonumber \\ && - m_{J_1}^{2} \tilde{J}^{\dagger}_1 \tilde{J}_1 - m_{J_{i}}^{2} \tilde{J}^{\dagger}_{i}\tilde{J}_{i} - m_{Q_{i}}^{2} \tilde{Q}^{ \dagger}_{i} \tilde{Q}_{i} 
-m_{ \chi}^{2} \chi^{ \dagger} \chi \nonumber \\ &&-  m_{ \rho}^{2} \rho^{ \dagger} \rho - m^2_{\chi'}\chi^{{}_{\prime}\dagger}\chi^{{}_{\prime}}-
m^2_{\rho^{{}_{\prime}}}\rho^{{}_{\prime}\dagger}\rho^{{}_{\prime}} +b_{\rho} \rho^{a}\rho^{{}_{\prime} a} \nonumber \\ && + b_{\chi}\chi^{a}\chi^{{}_{\prime} a} +
\tilde{Q}_{1}A^d_{1\alpha} \rho^{{}_{\prime}} \tilde{d}^{c}_{\alpha} + \tilde{Q}_{1} B^J_{11}\chi^{{}_{\prime}} \tilde{J}^{c}_1 \nonumber \\ && + \tilde{Q}_{i} A^u_{i\alpha} \rho \tilde{u}^{c}_{\alpha} + \tilde{Q}_{i} B^J_{ij}  \chi \tilde{J}^{c}_{j}. 
\label{softsusy}
\end{eqnarray}
The parameters in the bilinear terms on scalar fields have mass dimension two. Trilinear terms in the soft breaking lagrangean, bilinear terms in superpotential and gaugino mass terms all have parameters with mass dimension one. All the other parameters are dimensionless.

Considering the spontaneous breaking of the gauge symmetries, by supposing that $\langle \chi \rangle\,,\,\langle \chi^{\prime}\rangle >> \langle \rho \rangle\,,\,\langle \rho^{\prime}\rangle$, we get the following breaking sequence,
$$ 
{\rm SU(3)}_L \otimes {\rm
U(1)}_X \stackrel{ \langle \chi \rangle\,,\,\langle \chi^{\prime}\rangle} \Longrightarrow {\rm
SU(2)}_L\otimes{\rm U(1)}_Y \stackrel{\langle \rho \rangle\,,\,\langle \rho^{\prime}\rangle}
\Longrightarrow {\rm U(1)}_{\rm QED}.
$$ 
This spontaneous symmetry breakdown is appropriate to give masses to the gauge bosons, $V^{\pm}\,,\, U^{\pm \pm}\,,\, W^{\pm}\,,\, Z^{\prime}\,,\, Z$,  which are encoded in the following expressions,
\begin{eqnarray}
&&M_{{}_{W^{\pm}}}^2=\frac{g^2}{4}\left(v_{{}_{\rho}}^{2}+v_{{}_{\rho^{{}_{\prime}}}}^2\right),\,\,\,\,\,\,\,\,\,\,\,\,\,\,\,\,\,\,\,\,\,\,\,\,\,\,\,\,\,M_{{}_{Z}}^{2}=\frac{g^2}{4}\frac{\left(1+4t^2\right)}{\left(1+3t^2\right)}\left(v_{{}_{\rho}}^{2}+v_{{}_{\rho^{{}_{\prime}}}}^2\right), 
\nonumber \\
&&M^2_{Z^{\prime}}=\frac{g^2}{3}\left(1+3t^2\right)\left(v_{{}_{\chi}}^{2}+v_{{}_{\chi^{{}_{\prime}}}}^2\right),\,\,\,\,\,\,\, M_{{}_{U^{\pm \pm}}}^2=\frac{g^2}{4}\left(v_{{}_{\rho}}^{2}+v_{{}_{\rho^{{}_{\prime}}}}^2+v_{{}_{\chi}}^{2}+v_{{}_{\chi^{{}_{\prime}}}}^2\right)\nonumber \\
&&M_{{}_{V^{\pm}}}^2=\frac{g^2}{4}\left(v_{{}_{\chi}}^{2}+v_{{}_{\chi^{{}_{\prime}}}}^2\right),
\label{gbmass}
\end{eqnarray}
where we have denoted $\langle \chi \rangle=v_\chi$, $\langle \chi^{\prime} \rangle=v_{\chi^{\prime}}$, $\langle \rho \rangle=v_\rho$, $\langle \rho^{\prime} \rangle=v_{\rho^{\prime}}$,  $t=\frac{g_{N}}{g}$, with  $g_N$ being  the coupling constant associated to the gauge group $U(1)_N$  and $g$ is the gauge coupling for the $SU(3)_L$ gauge group (and also for the SM $SU(2)_L$, which is embedded in it).  As it should be, one of the gauge bosons remains massless, the photon, $A_\mu$. 

The masses of the charged leptons are obtained strictly from effective operators in the last term of the superpotential, Eq.~(\ref{E.O}),  
\begin{equation} \label{massalepton}
m_{\ell}= \frac{k_{l}}{2\Lambda}v_{\rho^{\prime}}v_{\chi^{\prime}}.
\end{equation}

Regarding the quark masses, the superpotential in Eq.~(\ref{firstpart}) along with the first two terms in Eq.~(\ref{E.O}) provide the following mass matrices for the up-type quarks in the basis $(u_1\,,\,u_2\,,\,u_3)$,
\be
M_{u}=\left(\begin{array}{ccc} 
						-\frac{k_{u{}_{11}}}{\Lambda} v_{\rho}v_{\chi} & -\frac{k_{u{}_{12}}}{\Lambda} v_{\rho}v_{\chi} & -\frac{k_{u{}_{13}}}{\Lambda} v_{\rho}v_{\chi} \\
						\lambda^{\prime \prime}_{{}_{21}} v_{\rho}		 & \lambda^{\prime \prime}_{{}_{22}} v_{\rho}     & \lambda^{\prime \prime}_{{}_{23}} v_{\rho}     \\
						\lambda^{\prime \prime}_{{}_{31}} v_{\rho}		 & \lambda^{\prime \prime}_{{}_{32}} v_{\rho}     & \lambda^{\prime \prime}_{{}_{33}} v_{\rho}     \\
\end{array}\right),
\ee
and for the down-type quarks in the basis $(d_1\,,\,d_2\,,\, d_3)$,
\be
M_{d}=\left(\begin{array}{ccc} 
						 \lambda^{\prime}_{{}_{11}} v_{\rho^{\prime}}                    & \lambda^{\prime}_{{}_{12}} v_{\rho^{\prime}}                    & \lambda^{\prime }_{{}_{13}} v_{\rho^{\prime}}  \\
						 \frac{k_{d{}_{21}}}{\Lambda} v_{\rho^{\prime}}v_{\chi^{\prime}} & \frac{k_{d{}_{22}}}{\Lambda} v_{\rho^{\prime}}v_{\chi^{\prime}} & \frac{k_{d{}_{23}}}{\Lambda} v_{\rho^{\prime}}v_{\chi^{\prime}} \\
						 \frac{k_{d{}_{31}}}{\Lambda} v_{\rho^{\prime}}v_{\chi^{\prime}} & \frac{k_{d{}_{32}}}{\Lambda} v_{\rho^{\prime}}v_{\chi^{\prime}} & \frac{k_{d{}_{33}}}{\Lambda} v_{\rho^{\prime}}v_{\chi^{\prime}}  \\
\end{array}\right).
\ee

We can naturally assign the values of $\Lambda$, $v_\chi$  and $v_{\chi^{\prime}}$  around TeV scale, while $v_\rho$  and $v_{\rho^{\prime}}$ lie in the electroweak scale and obey the bound $v^2_\rho + v^2_{\rho^{\prime}}=246^2$ GeV$^2$. Thus, for typical values of the Yukawa couplings, we can easily obtain the observed masses for all standard charged fermions.

\section{Scalar Sector}

In supersymmetric models the scalar sector receives contributions from three different sources that adds up to form the  scalar potential. These contributions are,

\be
V_{F}= \displaystyle\sum_{i} \Big| \frac{\partial \hat{f}}{\partial S_{i}}\Big|^{2}_{{}_{\hat{S}=S}},
\ee

\be
V_{D}= \frac{1}{2}\displaystyle\sum_{\alpha A}\left( \displaystyle\sum_{i} S_{i}^{{}_{\dagger}}g_{\alpha}t_{\alpha A }S_{i}\right)^{2},
\ee
and ,
\be
V_{soft}= m_{{}_{1}}^{2}\rho^{{}_{\dagger}}\rho + m_{{}_{2}}^{2}\rho^{{}_{\prime}{}_{\dagger}}\rho^{{}_{\prime}} + m_{{}_{3}}^{2}\chi^{{}_{\dagger}}\chi + m_{{}_{4}}^{2}\chi^{{}_{\prime}{}_{\dagger}}\chi^{{}_{\prime}} - b_{{}_{\rho}}\delta_{ab}\rho^{a}\rho^{{}_{\prime}b} - b_{{}_{\chi}}\delta_{ab}\chi^{a}\chi^{{}_{\prime}b},
\ee
where the summation index $i$ runs over all scalars, $\alpha$ runs through the different symmetry groups and $A$ through the group generators.

Working out the indices, we have,
\be
V_{F}= \mu_{{}_{\rho}}^{2}\big|\rho\big|^{2} + \mu_{{}_{\rho}}^{2}\big|\rho^{{}_{\prime}}\big|^{2} + \mu_{{}_{\chi}}^{2}\big|\chi\big|^{2} + \mu_{{}_{\chi}}^{2}\big|\chi^{{}_{\prime}}\big|^{2}
\ee
and
\be
V_{D}= \frac{g^{2}}{2}\left( \rho^{{}_{\dagger}}t_{A}\rho - \rho^{{}_{\prime}{}_{\dagger}}t^{\ast}_{A}\rho^{{}_{\prime}} + \chi^{{}_{\dagger}}t_{A}\chi - \chi^{{}_{\prime}{}_{\dagger}}t^{\ast}_{A}\chi^{{}_{\prime}}\right)^{2} + \frac{g_{N}^{2}}{2}\left( \rho^{{}_{\dagger}}\rho - \rho^{{}_{\prime}{}_{\dagger}}\rho^{{}_{\prime}} - \chi^{{}_{\dagger}}\chi + \chi^{{}_{\prime}{}_{\dagger}}\chi^{{}_{\prime}}\right)^{2}.
\ee

The scalar potential is the sum of the above three contributions,
\be \label{potential}
V= V_{F}+V_{D}+V_{soft}.
\ee

By performing the usual shift on the neutral scalars displaced by their respective VEVs,
\begin{eqnarray}
\rho^0,  \rho^{\prime 0}, \chi^0 , \chi^{\prime 0} \rightarrow \frac{1}{\sqrt{2}} (v_{ \rho,\rho^{\prime }, \chi , \chi^{\prime}} + R_{ \rho,\rho^{\prime }, \chi , \chi^{\prime}} + iI_{ \rho,\rho^{\prime }, \chi , \chi^{\prime}}),
\label{scalarshift}
\end{eqnarray}
the set of minimum conditions are given by,
\bea
\left< \dfrac{\partial V}{\partial \rho^{{}_{0}}}\right>_{0} &=& g^2\left(2v_{\rho}^2 - 2 v_{\rho^{{}_{\prime}}}^2 - v_{\chi}^2 + v_{\chi^{{}_{\prime}}}^2\right) + 6\left(v_{\rho}^2 - v_{\rho^{{}_{\prime}}}^2 -v_{\chi}^2 + v_{\chi^{{}_{\prime}}}^2 \right)g_{N}^2 + 12 m_{{}_{1}}^2 + 12 \mu_{{}_{\rho}}^2 - 12 \frac{v_{\rho^{{}_{\prime}}}}{v_{\rho}} b_{{}_{\rho}}=0\,, \nonumber \\
\left< \dfrac{\partial V}{\partial \rho^{{}_{\prime 0}}}\right>_{0} &=& g^2\left(-2v_{\rho}^2 + 2 v_{\rho^{{}_{\prime}}}^2+ v_{\chi}^2 - v_{\chi^{{}_{\prime}}}^2\right) - 6\left(v_{\rho}^2 - v_{\rho^{{}_{\prime}}}^2 -v_{\chi}^2 + v_{\chi^{{}_{\prime}}}^2 \right)g_{N}^2 + 12 m_{{}_{2}}^2 + 12 \mu_{{}_{\rho}}^2 - 12 \frac{v_{\rho}}{v_{\rho^{{}_{\prime}}}} b_{{}_{\rho}}=0\,, \nonumber \\
\left< \dfrac{\partial V}{\partial \chi^{{}_{0}}}\right>_{0} &=& g^2\left(-v_{\rho}^2 + v_{\rho^{{}_{\prime}}}^2+ 2v_{\chi}^2 - 2v_{\chi^{{}_{\prime}}}^2\right) - 6\left(v_{\rho}^2 - v_{\rho^{{}_{\prime}}}^2 -v_{\chi}^2 + v_{\chi^{{}_{\prime}}}^2 \right)g_{N}^2 + 12 m_{{}_{3}}^2 + 12 \mu_{{}_{\chi}}^2 - 12 \frac{v_{\chi^{{}_{\prime}}}}{v_{\chi}} b_{{}_{\chi}}=0\,, \nonumber \\
\left< \dfrac{\partial V}{\partial \chi^{{}_{\prime 0}}}\right>_{0} &=& g^2\left(v_{\rho}^2 - v_{\rho^{{}_{\prime}}}^2 - 2v_{\chi}^2 + 2v_{\chi^{{}_{\prime}}}^2\right)+6\left(v_{\rho}^2 - v_{\rho^{{}_{\prime}}}^2 -v_{\chi}^2 + v_{\chi^{{}_{\prime}}}^2 \right)g_{N}^2 + 12 m_{{}_{4}}^2 + 12 \mu_{{}_{\chi}}^2 - 12 \frac{v_{\chi}}{v_{\chi^{{}_{\prime}}}} b_{{}_{\chi}}=0\,. \nonumber \\
\label{vinculos}
\eea

With this set of constraint  equations, we are able to obtain the texture of the scalar mass matrices in the model. 

We start with the CP-odd scalars, that lead to two $2\times 2$  mass matrices. The first one, in the basis $(I_{\chi},I_{\chi^{{}_{\prime}}})^T$, takes the form, 
\be
\left(\begin{array}{cc}
										  \frac{v_{\chi^{{}_{\prime}}}b_{{}_{\chi}}}{2v_{{}_{\chi}}} & \frac{b_{{}_{\chi}}}{2} \\
											\frac{b_{{}_{\chi}}}{2}                                    & \frac{v_{\chi}b_{{}_{\chi}}}{2v_{\chi^{{}_{\prime}}}}
										 \end{array}\right)\label{Ichi},										 
\ee
while the second one, in the basis $(I_{\rho},I_{\rho^{{}_{\prime}}})^T$, takes the form, 
\be
\left(\begin{array}{cc}
										  \frac{v_{\rho^{{}_{\prime}}}b_{{}_{\rho}}}{2v_{{}_{\rho}}} & \frac{b_{{}_{\rho}}}{2} \\
											\frac{b_{{}_{\rho}}}{2}                                    & \frac{v_{\rho}b_{{}_{\rho}}}{2v_{\rho^{{}_{\prime}}}}
										 \end{array}\right)\label{Irho}.
\ee

Both matrices  in Eqs. (\ref{Ichi}) and (\ref{Irho}) have the same pattern and are easily diagonalized, providing the following eigenvalues,
\bea
M_{A}^2 &=& \frac{\left(v_{\rho^{{}_{\prime}}}^2 +v_{\rho}^2 \right)b_{{}_{\rho}}}{v_{\rho}v_{\rho^{{}_{\prime}}}},
\nonumber \\
M_{A^{\prime}}^2 &=& \frac{\left(v_{\chi^{{}_{\prime}}}^2 +v_{\chi}^2 \right)b_{{}_{\chi}}}{v_{\chi}v_{\chi^{{}_{\prime}}}},
\nonumber \\
M_{G}^2 &=& M_{G^{\prime}}^2 = 0,
\label{pseudomass}
\eea
whose eigenstates are respectively given by,
\bea
A &=& \frac{v_{\rho^{{}_{\prime}}}}{\sqrt{v_{\rho}^2+v_{\rho^{{}_{\prime}}}^2}}I_{\rho} + \frac{v_{\rho}}{\sqrt{v_{\rho}^2+v_{\rho^{{}_{\prime}}}^2}}I_{\rho^{{}_{\prime}}}, \nonumber \\
A^{\prime} &=& \frac{v_{\chi^{{}_{\prime}}}}{\sqrt{v_{\chi}^2+v_{\chi^{{}_{\prime}}}^2}}I_{\chi} + \frac{v_{\chi}}{\sqrt{v_{\chi}^2+v_{\chi^{{}_{\prime}}}^2}}I_{\chi^{{}_{\prime}}}, \nonumber \\
G &=& -\frac{v_{\rho}}{\sqrt{v_{\rho}^2+v_{\rho^{{}_{\prime}}}^2}}I_{\rho} + \frac{v_{\rho^{{}_{\prime}}}}{\sqrt{v_{\rho}^2+v_{\rho^{{}_{\prime}}}^2}}I_{\rho^{{}_{\prime}}}, \nonumber \\
G^{\prime}&=& -\frac{v_{\chi}}{\sqrt{v_{\chi}^2+v_{\chi^{{}_{\prime}}}^2}}I_{\chi} + \frac{v_{\chi^{{}_{\prime}}}}{\sqrt{v_{\chi}^2+v_{\chi^{{}_{\prime}}}^2}}I_{\chi^{{}_{\prime}}},
\label{pseudostates}
\eea
where  $A$ and $A^\prime$ are the massive CP-odd states and $G$ and $G^\prime$ are the Goldstone bosons eaten by the neutral gauge bosons $Z$ and $Z^{\prime}$.

For the singly-charged scalars we also have two $2 \times 2$ mass matrix. The first one, in the basis $\left( \rho^{+} , \rho^{{}_{\prime}+ }\right)^T$, takes the form,
\be
\left(
\begin{array}{cc}
 \frac{g^2 v_{\rho^{{}_{\prime}}}^2}{4}+\frac{b_\rho v_{\rho_{{}_{\prime}}}}{v_{\rho}} & -(\frac{1}{4}
   v_{\rho} v_{\rho^{{}_{\prime}}} g^2+b_\rho) \\
 -(\frac{1}{4} v_{\rho} v_{\rho^{{}_{\prime}}} g^2+b_\rho) & \frac{g^2 v_{\rho}^2}{4}+\frac{b_\rho v_{\rho}}{v_{\rho^{{}_{\prime}}}}
\end{array}
\label{rho+}
\right),
\ee
while the second one, in the basis  $\left( \chi^{+} , \chi^{{}_{\prime}+}\right)^T$, is,
\be
\left(
\begin{array}{cc}
 \frac{g^2 v_{\chi^{{}_{\prime}}}^2}{4}+\frac{b_\chi v_{\chi_{{}_{\prime}}}}{v_{\chi}} & -(\frac{1}{4}
   v_{\chi} v_{\chi^{{}_{\prime}}} g^2+b_\chi) \\
 -(\frac{1}{4} v_{\chi} v_{\chi^{{}_{\prime}}} g^2+b_\chi) & \frac{g^2 v_{\chi}^2}{4}+\frac{b_\chi v_{\chi}}{v_{\chi^{{}_{\prime}}}}
\end{array}
\label{chi+}
\right).
\ee

These two matrices have eigenvalues,
\bea
M_{H^{+}}^2 &=& M_{A}^2+M_{W}^2, \nonumber \\
M_{H^{{}_{\prime}+}}^2 &=& M_{A^{{}_{\prime}}}^2+M_{V}^2, \label{MH+}\\
M_{G^{+}}^2 &=& M_{G^{{}_{\prime}+}}^2 = 0 \nonumber
\label{chargedsmass}
\eea
with the respective eigenvectors,
\bea
H^{+} &=& -\frac{v_{\rho^{{}_{\prime}}}}{\sqrt{v_{\rho}^2 + v_{\rho^{{}_{\prime}}}^2}} \rho^{+} +\frac{v_{\rho}}{\sqrt{v_{\rho}^2 + v_{\rho^{{}_{\prime}}}^2}} \rho^{{}_{\prime}+}, \nonumber \\
H^{{}_{\prime}+} &=& -\frac{v_{\chi^{{}_{\prime}}}}{\sqrt{v_{\chi}^2 + v_{\chi^{{}_{\prime}}}^2}} \chi^{+} +\frac{v_{\chi}}{\sqrt{v_{\chi}^2 + v_{\chi^{{}_{\prime}}}^2}} \chi^{{}_{\prime}+}, \nonumber \\
G^{+} &=& \frac{v_{\rho}}{\sqrt{v_{\rho}^2 + v_{\rho^{{}_{\prime}}}^2}} \rho^{+} +\frac{v_{\rho^{{}_{\prime}}}}{\sqrt{v_{\rho}^2 + v_{\rho^{{}_{\prime}}}^2}} \rho^{{}_{\prime}+},  \\
G^{{}_{\prime}+} &=& \frac{v_{\chi}}{\sqrt{v_{\chi}^2 + v_{\chi^{{}_{\prime}}}^2}} \chi^{+} +\frac{v_{\chi^{{}_{\prime}}}}{\sqrt{v_{\chi}^2 + v_{\chi^{{}_{\prime}}}^2}} \chi^{{}_{\prime}+}, \nonumber
\label{chargedsstates}
\eea
where $M_{V}$ and $M_W$ stand for the gauge boson masses in Eq.~(\ref{gbmass}). As can be seen, there are two Goldstone bosons, those eaten by the two singly charged gauge bosons. 

For the doubly-charged scalars, we are going to have a $4\times 4$ mass matrix which, in the  basis $\left( \rho^{++} , \rho^{{}_{\prime}++}, \chi^{++} , \chi^{\prime++}\right)^T$, takes the form,
\be
\left(\begin{array}{cccc}
									\frac{g^2v_{\rho} \left(v_{\rho^{{}_{\prime}}}^2+v_{\chi}^2-v_{\chi^{{}_{\prime}}}^2\right) +4v_{\rho^{{}_{\prime}}} b_\rho}{4 v_{\rho}} & -\frac{g^2v_{\rho}v_{\rho^{{}_{\prime}}}}{4} -b_\rho& \frac{g^2 v_{\rho}v_{\chi}}{4}& -\frac{g^2 v_{\rho} v_{\chi^{{}_{\prime}}}}{4}  \\
									-\frac{g^2v_{\rho}v_{\rho^{{}_{\prime}}}}{4}-b_\rho & \frac{g^2v_{\rho^{{}_{\prime}}} \left(v_{\rho}^2-v_{\chi}^2+v_{\chi^{{}_{\prime}}}^2\right) +4v_{\rho} b_\rho}{4 v_{\rho^{{}_{\prime}}}}   & -\frac{g^2 v_{\rho^{{}_{\prime}}}v_{\chi}}{4} & \frac{g^2 v_{\rho^{{}_{\prime}}}v_{\chi^{{}_{\prime}}}}{4} \\
										\frac{g^2 v_{\rho}v_{\chi}}{4}  & -\frac{g^2v_{\rho^{{}_{\prime}}}v_{\chi} }{4} & \frac{g^2v_{\chi}\left(v_{\rho}^2-v_{\rho^{{}_{\prime}}}^2+v_{\chi^{{}_{\prime}}}^2\right) +4v_{\chi^{{}_{\prime}}}b_\chi}{4 v_{\chi}} &-\frac{g^2v_{\chi} v_{\chi^{{}_{\prime}}}}{4} -b_\chi \\
									-\frac{g^2 v_{\rho}v_{\chi^{{}_{\prime}}}}{4} & \frac{ g^2 v_{\rho^{{}_{\prime}}}v_{\chi^{{}_{\prime}}}}{4} &-\frac{g^2v_{\chi}v_{\chi^{{}_{\prime}}}}{4} -b_\chi & \frac{g^2v_{\chi^{{}_{\prime}}}\left(-v_{\rho}^2+v_{\rho^{{}_{\prime}}}^2+v_{\chi}^2\right)+4v_{\chi} b_\chi}{4v_{\chi^{{}_{\prime}}}}
\end{array}\right).
\label{M++}
\ee

To obtain analytical eigenvalues and eigenstates for this matrix is a somewhat cumbersome task that we will not follow. However, its determinant is equal to zero, which provides (after a thorough numerical analysis of all the eigenvalues) only one null eigenvalue that will be the Goldstone eaten by the gauge boson $U^{++}$. Later, in the next section, when we have specified some of the model parameters, we will present the range of mass values for the lightest doubly-charged scalar, which will be around some few hundreds of GeV.

Let us finally focus on the CP-even scalars. Considering the  basis $\left(R_{\rho},R_{\rho^{{}_{\prime}}},R_{\chi},R_{\chi^{{}_{\prime}}}\right)^T$, its mass matrix takes the following form,
\be
{\scriptsize
\left(
\begin{array}{cccc}
 \left(\frac{g^2}{3}+g_{N}^2\right)v^2 \sin^2 \beta+\frac{b_{\rho}}{ \tan \beta} & -\left(\frac{g^2}{3}+g_{N}^2\right)v^2\cos \beta \sin \beta - b_{\rho} & -\left(\frac{g^2}{6}+g_{N}^2\right)v v_{\chi} \sin \beta & \left(\frac{g^2}{6}+g_{N}^2\right)v v_{\chi^{\prime}} \sin \beta \\
-\left(\frac{g^2}{3}+g_{N}^2\right)v^2\cos \beta \sin \beta - b_{\rho} & \left(\frac{g^2}{3}+g_{N}^2\right)v^2 \cos^2 \beta+b_{\rho}\tan \beta  & \left(\frac{g^2}{6}+g_{N}^2\right)vv_{\chi}\cos \beta & -\left(\frac{g^2}{6}+g_{N}^2\right)v v_{\chi^{\prime}} \cos \beta\\
-\left(\frac{g^2}{6}+g_{N}^2\right)v v_{\chi}\sin \beta &  \left(\frac{g^2}{6}+g_{N}^2\right)v v_{\chi}\cos \beta &  \left(\frac{g^2}{3}+g_{N}^2\right)v_{\chi}^2+b_{\chi}\frac{v_{\chi^{\prime}}}{v_{\chi}} &  -\left(\frac{g^2}{3}+g_{N}^2\right)v_{\chi}v_{\chi^{\prime}} - b_{\chi} \\
\left(\frac{g^2}{6}+g_{N}^2\right)v v_{\chi^{\prime}}\sin \beta & -\left(\frac{g^2}{6}+g_{N}^2\right)v v_{\chi^{\prime}}\cos \beta &   -\left(\frac{g^2}{3}+g_{N}^2\right)v_{\chi}v_{\chi^{\prime}} - b_{\chi} & \left(\frac{g^2}{3}+g_{N}^2\right)v_{\chi_{\prime}}^2+b_{\chi}\frac{v_{\chi}}{v_{\chi^{\prime}}}
\end{array}
\right),} \label{cpeven}
\ee
where 
\be \label{tanbeta}
\tan{\beta}=\frac{v_{\rho}}{v_{\rho^{{}_{\prime}}}},
\ee
and $v$ is the SM electroweak symmetry breaking scale given by $v^2=v_\rho^2 + v_\rho^{\prime 2}$.
From this mass matrix we are going to have four massive scalars, from which the lightest one reproduces the properties of the SM Higgs, which we assume as the scalar boson recently found in the LHC. In order to guarantee that such a scalar plays the role of a Higgs boson, we are going to demand throughout this work that its eigenstate is at least $95\%$ composed of the real part of $\rho^0$. This is mandatory to certify that such CP-even scalar behaves very much like the SM Higgs boson, since the two first components of the triplet $\rho$ mimics the SM Higgs doublet in the context of minimal 331 model. Besides, this choice assures us that its branching ratios to SM particles are basically the same computed in Ref.~\cite{2fotons}, except for some minor corrections coming from the extra particles in the SUSY spectrum (an analysis we are going to pursue somewhere else).
In the next section we study the behavior of the lightest scalars of the model.

\section{Higgs Phenomenology and Scalar Sector}

To assess the capability of this model to reproduce the results of the ATLAS and CMS collaboration~\cite{LHC}, it will be necessary to define the parameter space responsible for the Higgs mass. As can be seen from the mass matrices of all  scalars of the model, there are five free parameters $\left(\beta , v_{\chi}, v_{\chi^{\prime}}, b_{\rho}, b_{\chi}\right)$ that define their eigenvalues. Before proceeding further, it makes necessary to call the attention to  some features of the RSUSY331 model: First of all, charged lepton masses, through Eq.~(\ref{massalepton}), impose restrictions to $v_{\rho^{\prime}}$ and $v_{\chi^{\prime}}$  in order  to avoid entering some nonperturbative  regime where  $k_l>\sqrt{4\pi}$, a worry only justified for the case of the tau lepton;
Secondly,  we establish that  $v_{\chi}^2+v_{\chi^{\prime}}^2=v_{331}^2$, where $v_{331}$ is the energy scale  characteristic of the 331 spontaneous symmetry breakdown. It is important to notice that $v_{331}$ should not exceed the cutoff scale ($\Lambda$), otherwise the theory would fall in the non-perturbative regime concerning the $U(1)_N$ gauge coupling; Third, the soft SUSY breaking parameters, $b_{\rho}$ and $b_{\chi}$, have mass dimension two and we can think of it as a product of two mass scales of  order of SUSY breaking, that is, $b_{\rho},b_{\chi}\sim M_{SUSY}^2$. Assuming that the SUSY breaking scale is, roughly speaking, of the order of few TeV, the range of these two paramaters can be set, without loss of generality, to
\be 
10^5 \, \, \text{GeV}^2 < b_{\rho},b_{\chi} \leq 10^6 \, \, \text{GeV}^2\,;
\label{range}
\ee
Lastly, we will require that the squared masses of all scalars of the model  be real and positive. This condition restricts $ b_{\rho}$ and $b_{\chi}$ in a way that tachyons are not present in the scalar spectrum, a problem that had to be circumvented in Ref.~\cite{susylong} at the expense of having two massless doubly charged scalars in the spectrum, something we definitely do not desire to happen in our model~\footnote{We guess that such differences between these models may be related to the different R-parity symmetries chosen, besides some terms in the SUSY soft breaking lagrangian as well as in the superpotential.}.  The choice of the range of parameters  made in Eq.~(\ref{range}) already satisfies this condition. For the other parameters of the model, we consider they vary inside the following range of values,
\bea \label{parameter}
& 1000 \, \, \text{GeV} \leq v_{331} \leq \, \, 4500 \, \, \text{GeV}, & \\
& 0.1 \leq k_{\tau} \leq 0.9, & \\
& 0 < \beta < \dfrac{\pi}{2}. &
\eea

\begin{figure}[!hb]
\centering
\includegraphics[scale=0.75]{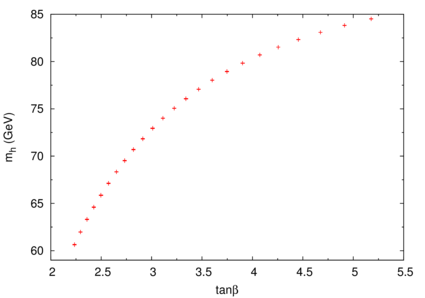}
\caption{{\footnotesize Values of $\tan \beta$ compatible with tree level Higgs mass $m_h>60$. As can be seen, there is a maximum value of $m_h$ given the assumptions made, where $tan \beta\approx 5$.} }
\label{fig:1}
\end{figure}

\begin{figure}[!h]
$$\begin{array}{c}
\begin{minipage}[b]{0.45\linewidth}
\includegraphics[scale=0.5]{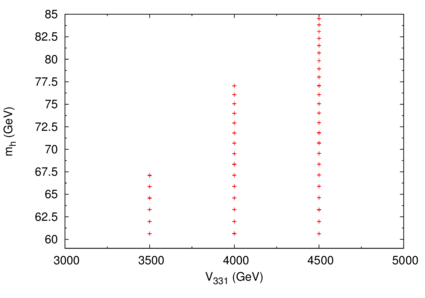}
\end{minipage} \hfill
\begin{minipage}[b]{0.45\linewidth}
\includegraphics[scale=0.5]{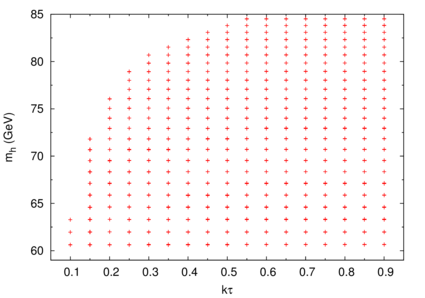}
\end{minipage} \\
\begin{minipage}[b]{0.45\linewidth}
\includegraphics[scale=0.5]{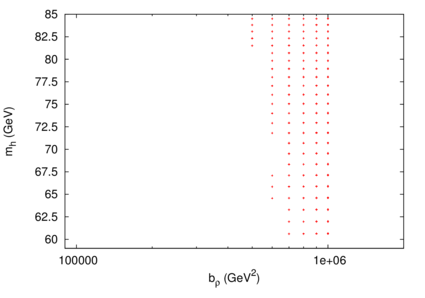}
\end{minipage} \hfill
\begin{minipage}[b]{0.45\linewidth}
\includegraphics[scale=0.5]{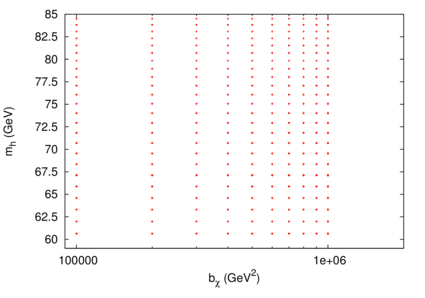}
\end{minipage}
\end{array}$$
\caption{{\footnotesize Behavior of the free parameters compared with the higgs mass $m_h$. The 331 breaking scale, $v_{331}$, must have the highest possible value to achieve the highest tree level mass for the Higgs boson (top left); The $\tau$ lepton Yukawa coupling, $k_\tau$, must be greater than 0.5 to reach the maximum value of the Higgs mass (top right); Although $b_\rho$ must be greater than $5 \times 10^5$ GeV$^2$ to have the highest higgs mass (bottom left), $b_\chi$ is insensitive to such mass limit (bottom right).}}
\label{fig:2}\end{figure}

In FIGs.\ref{fig:1} and \ref{fig:2} we  show the behavior of the lightest scalar mass, $m_h$,  at tree level  with the  free parameters of the model.  

Perceive that we have the upper bound of $m_h < 90$GeV at tree level. This recovers, in part,   the behavior of the Higgs  in the MSSM and is consistent with previous estimate in an enlarged version of this model~\cite{DuongMa}. Thus, we conclude that this scalar should play the role of the Higgs in the RSUSY331 model. Hence, a Higgs with mass of 125~GeV, as measured by CMS and ATLAS, demands substantial contribution from radiative corrections, as is the case for the MSSM. We stress that classical conditions for the stability of the potential and absence of tachyons in our model refrain the  mass of the Higgs in the RSUSY331 model from going beyond  $90$ GeV at tree level. In this sense, it turns imperative  to analyze the main loop contributions to the mass of the Higgs in the RSUSY31 model. As is widely known, the stop gives the main  contribution to the Higgs mass.  In our case, the  stop loop correction to the Higgs mass will be calculated using the effective potential approach~\cite{potencialefetivosusy}. For this we first have to evaluate the mass matrix of the stops. This is given by,
\be
\left(
\begin{array}{cc}
m_{t}^2 + m_{Q_{3}}^2 +\frac{1}{3}\left(\dfrac{g^2}{2}+g_{N}^2 \right) \Delta v_\chi & m_{t}X_{t} \\
m_{t}X_{t} & m_{t}^2+ m_{u_{3}}^2 + \frac{2}{3}g_{N}^2\Delta v_\chi
\end{array}
\label{MMstops}
\right).
\ee
where D-Terms which contributes minimaly where neglected. Its eigenvalues are, 
\bea
m_{\tilde{t}_{1}}^2 & = & m_{t}^2 +\frac{1}{2}\left( m_{Q_{3}}^2 +\frac{1}{3}\left(\dfrac{g^2}{2}+g_{N}^2\right)\Delta v_\chi \right) +\frac{1}{2}\left( m_{u_{3}}^2 +\frac{2}{3}g_{N}^2\Delta v_\chi \right) \nonumber \\
&  & + \frac{1}{2}\sqrt{\left(\left( m_{Q_{3}}^2 +\frac{1}{3}\left(\dfrac{g^2}{2}+g_{N}^2\right)\Delta v_\chi \right)-\left( m_{u_{3}}^2 +\frac{2}{3}g_{N}^2\Delta v_\chi \right)\right)^2+4m^2_{t}X^2_{t}}\,, \nonumber \\
m_{\tilde{t}_{2}}^2 & = & m_{t}^2 +\frac{1}{2}\left( m_{Q_{3}}^2 +\frac{1}{3}\left(\dfrac{g^2}{2}+g_{N}^2\right)\Delta v_\chi \right) +\frac{1}{2}\left( m_{u_{3}}^2 +\frac{2}{3}g_{N}^2\Delta v_\chi \right) \\ 
            &  & - \frac{1}{2}\sqrt{\left(\left( m_{Q_{3}}^2 +\frac{1}{3}\left(\dfrac{g^2}{2}+g_{N}^2\right)\Delta v_\chi \right)-\left( m_{u_{3}}^2 +\frac{2}{3}g_{N}^2\Delta v_\chi \right)\right)^2+4m^2_{t}X^2_{t}} \,,\nonumber
\label{mstops}
\eea
where $m_{Q_3}$ and  $m_{u_3}$ are  soft SUSY  breaking terms given in Eq.~(\ref{softsusy}), $\Delta v_\chi = \frac{1}{2}\left(v_{\chi}^2 - v_{\chi^{{}_{\prime}}}^2\right)$ and $X_{t}= A_{t}+\mu_\rho \cot \beta$ is the mixing parameter between $\tilde{t}_{L}$ and $\tilde{t}_{R}$ (we have identified $A_t\equiv A^u_{33}$ from Eq.~(\ref{softsusy})). It is opportune to remark that Eq.~(\ref{mstops}) dictates that the lightest stop is $\tilde{t}_2$.

In a first approximation, the mass matrix of the CP even scalars given in Eq.~(\ref{cpeven}) gets one-loop corrections only in the entries $11$, $12$ and $22$, which are
\bea
 \delta M_{11} &=& \frac{3G_{F}m_{t}^4 \mathrm{cosec}^2 \beta \left( A_t\left(A_t + \mu_{\rho}\cot \beta\right)\ln(\frac{m_{\tilde{t}_1}^4}{m_{\tilde{t}_2}^4}) - (m_{\tilde{t}_1}^2 -m_{\tilde{t}_2}^2)\ln(\frac{m_{t}^4}{m_{\tilde{t}_1}^2 m_{\tilde{t}_2}^2})\right) }{2\sqrt{2}\pi^2 \left(m_{\tilde{t}_1}^2-m_{\tilde{t}_2}^2\right)} \nonumber \\ & &+ \frac{3 G_{F} m_{t}^4 A_{t}^2\left(A_t + \mu_{\rho}\cot \beta\right)^2 \mathrm{cosec}^2 \beta\left(2m_{\tilde{t}_1}^2 -2m_{\tilde{t}_2}^2 -\left(m_{\tilde{t}_1}^2+m_{\tilde{t}_2}^2\right)\ln(\frac{m_{\tilde{t}_1}^2}{m_{\tilde{t}_2}^2})\right)}{2\sqrt{2}\pi^2 \left(m_{\tilde{t}_1}^2-m_{\tilde{t}_2}^2\right)^3}, \nonumber \\
\eea
\bea
 \delta M_{12} &=& \frac{3 G_{F} m_{t}^4 \mu_{\rho}^2\left(A_t + \mu_{\rho}\cot \beta\right)^2  \mathrm{cosec}^2 \beta \ln(\frac{m_{\tilde{t}_1}^2}{m_{\tilde{t}_2}^2})}{2\sqrt{2}\pi^2 \left(m_{\tilde{t}_1}^2-m_{\tilde{t}_2}^2\right)} \nonumber \\  & & + \frac{3G_{F} m_{t}^4 A_{t}\mu_{\rho} \left(A_t + \mu_{\rho}\cot \beta\right)^2 \mathrm{cosec}^2 \beta\left(2m_{\tilde{t}_1}^2 -2m_{\tilde{t}_2}^2 -\left(m_{\tilde{t}_1}^2+m_{\tilde{t}_2}^2\right)\ln(\frac{m_{\tilde{t}_1}^2}{m_{\tilde{t}_2}^2})\right)}{2\sqrt{2}\pi^2 \left(m_{\tilde{t}_1}^2-m_{\tilde{t}_2}^2\right)^3}, \nonumber \\
\eea
\be
 \delta M_{22}= \frac{3 G_{F} m_{t}^4\mu_{\rho}^2\left(A_t + \mu_{\rho}\cot \beta \right)^2 \mathrm{cosec}^2 \beta\left(2m_{\tilde{t}_1}^2 -2m_{\tilde{t}_2}^2 -\left(m_{\tilde{t}_1}^2+m_{\tilde{t}_2}^2\right)\ln\left(\frac{m_{\tilde{t}_1}^2}{m_{\tilde{t}_2}^2}\right)\right)}{2\sqrt{2}\pi^2 \left(m_{\tilde{t}_1}^2-m_{\tilde{t}_2}^2\right)^3}, 
\ee

We will re-analyze the mass matrix in Eq.~(\ref{cpeven}) but now taking into account the above corrections. However, we select values of the free parameters that maximize the value of $m_h$ at tree level, as well as impose that the Higgs is more than $95\%$ composed of $\rho^0$, meaning that our analysis is now restricted to a  narrower range of values of the parameters that enter in the calculations, namely,
\bea 
& 2500 \, \, \text{TeV} \leq v_{331} \leq \, \, 4500 \, \, \text{TeV}, & \\
& 1 < \beta < 1.56, & \\
& k_{\tau}=0.5, & \\
& 10^5 \, \, \text{GeV}^2 < b_{\rho},b_{\chi} \leq 10^6 \, \, \text{GeV}^2 .
\label{parameter2}
\eea
We have also chosen the  soft breaking stop mass, $m_{Q_{3}}$ and $m_{u_{3}}$, between $100$~GeV and $1.5$~TeV, and the trilinear soft breaking parameter $-2 \, \, \text{TeV} < A_t <2 \, \, \text{TeV}$. 

\begin{figure}[!h]
\centering
\includegraphics[scale=0.75]{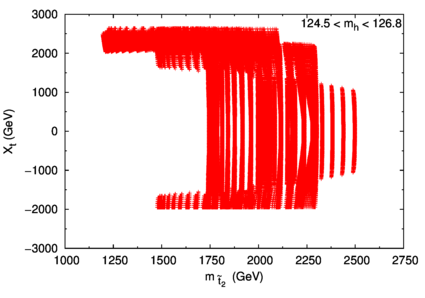}
\caption{{\footnotesize Stop mass and mixing necessary to obtain a Higgs mass in the region $124.5$~GeV$\leq m_{h} \leq 126.8$~GeV.  }}
\label{fig:3}
\end{figure}

On considering all these assumptions, we are able to present the stop mass and mixing necessary to fit the recent data reported by CMS and ATLAS experiments.  In FIG.~\ref{fig:3} we present the behavior of the lightest stop in function of their mixing. 

\begin{figure}[!h]
\begin{minipage}[b]{0.45\linewidth}
\includegraphics[scale=0.5]{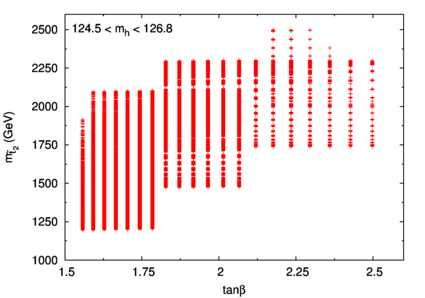}
\end{minipage} \hfill
\begin{minipage}[b]{0.45\linewidth}
\includegraphics[scale=0.5]{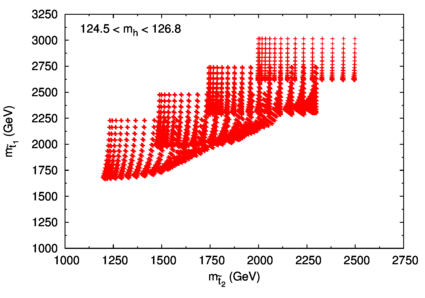}
\end{minipage}
\caption{Comparison of the lightest stop mass with $\tan \beta$ and the heaviest stop mass. (left) $\tan \beta$ must lie between 1.5 and 2.5 to be in agreement with the higgs mass constraint. (right) The heaviest stop must be heavier than $m_{{\tilde{t}_1}}\approx 1625$~GeV, for the smallest light stop mass $m_{{\tilde{t}_2}}\approx 1180$~GeV,.}
\label{fig:4}
\end{figure}

Differently from the MSSM case,  the lightest stop cannot be as light as $1180$~GeV. However, the stop mixing can be small and even negligible if the stop mass is higher than $1750$~GeV. If we focus on Eq.~(\ref{mstops}), we can see that, even if the soft breaking parameters were of the order of hundreds of GeV, we would have a stop mass above TeV because  the $\Delta v_\chi$ term will drive the stop masses to the TeV scale. Moreover, the stop soft mass parameters can be as low as $100$~GeV. This is interesting because it requires less   fine tuning compared to the MSSM~\cite{finetuning}.  For sake of completeness, in FIG.~\ref{fig:4} we present the behavior of the lightest stop mass in function of $\tan \beta$ and compare it with  the mass of the heaviest stop. 

\begin{figure}[!h]
$$\begin{array}{c}
\begin{minipage}[b]{0.45\linewidth}
\includegraphics[scale=0.5]{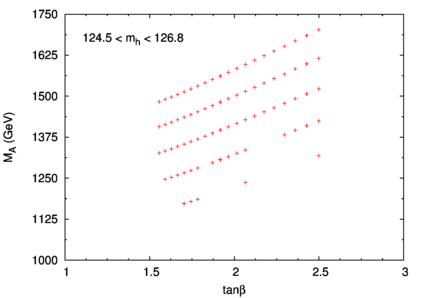}
\end{minipage} \hfill
\begin{minipage}[b]{0.45\linewidth}
\includegraphics[scale=0.5]{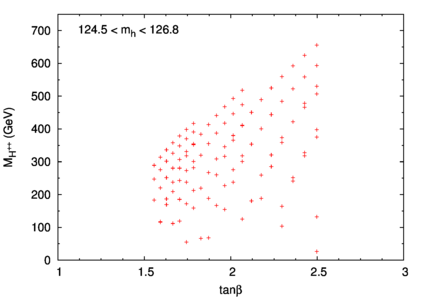}
\end{minipage} \\
\includegraphics[scale=0.5]{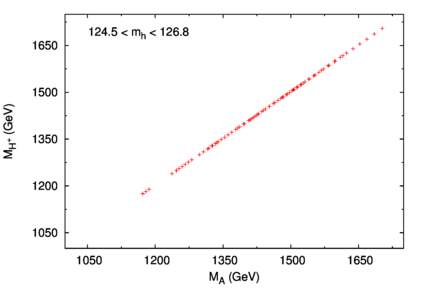}
\end{array}$$
\caption{{\footnotesize CP odd, singly and doubly charged scalars masses. (top left) The CP odd scalar mass must lie between $1125$ GeV and $1700$ GeV for 1.5$<\tan \beta <$2.5. (top right) For $\tan \beta$ in the same range, the doubly charged Higgs mass lies between $25$ GeV and $650$ GeV. (bottom) As is expected in a decoupled Higgs sector, the singly charged scalar is almost perfectly degenerate with the CP odd one as predicted by Eq. (\ref{MH+}).}}
\label{fig:5}
\end{figure}

On regarding the other scalars of the model, we present the behavior of the lightest singly and doubly  charged scalars and the CP odd one.  FIG.~\ref{fig:5} tells us that these scalars, with the exception of the doubly charged one, lie at the TeV scale. 
As a nice result, observe that the mass of the doubly charged scalar may be as low as $250$ GeV, which can be  probed in the LHC.

\section{Conclusions}

In this work we have developed the scalar sector of the recently proposed RSUSY331 model, concentrating on the mass of Higgs boson, which we enforced to almost match the SM one. We have shown that, similarly to the MSSM case, a Higgs with mass of 125 GeV requires robust  radiative corrections.  However, differently from the MSSM case, the radiative corrections require stops with mass at TeV scale, but with small mixing. Moreover, the soft breaking mass terms are free to be as light as hundreds of GeV. This is nice because we can avoid substantial fine tuning, differently from the MSSM.

 The model predicts that the lightest doubly charged scalar has mass at the electroweak scale which can be probed at the LHC. The remaining scalars of the model must have masses at the TeV scale. Most importantly, our framework naturally avoid tachyons or unwanted massless charged scalars in its spectrum compared to Ref.~\cite{susylong}. Finally, this SUSY version of the minimal 331 has a reduced scalar sector suitable to perform further phenomenological analysis and the lightest supersymmetric particle can be the dark matter candidate, not present in the non-supersymmetric version, an issue to be investigated somewhere else.
\acknowledgments
This work was supported by Conselho Nacional de Desenvolvimento Cient\'{i}fico e Tecnol\'ogico - CNPq (C.A.S.P. and P.S.R.S.) and Coordena\c c\~ao de Aperfei\c coamento de Pessoal de N\'{\i}vel Superior - CAPES (A.S. and J.G.F.Jr.).
%


\begin{thebibliography}{99}
%
\bibitem{LHC}
G. Aad \textit{et al}. (Atlas Collaboration), Phys. Lett.  {\bf B716},  (2012) 1;
S. Chatrchyan \textit{et al}. (CMS Collaboration), Phys. Lett. {\bf B716},  (2012) 30;
Moriond Proceeding contributions, \url{http://moriond.in2p3.fr/QCD/2013/MorQCD13Prog.html}.
%
\bibitem{neutrinos}
SNO Collaboration (Q. R. Ahmad {\it et al.}), Phys. Rev. Lett. {\bf 89}, (2002) 011301; ibidem, Phys. Rev. Lett. {\bf 89}, (2002) 011302;
Super-Kamiokande Collaboration (J. Hosaka {\it et al}.), Phys. Rev. D{\bf 74}, (2006) 032002;
KamLAND Collaboration (K. Eguchi {\it et al}.), Phys. Rev. Lett. {\bf 90}, (2003) 021802;
K2K Collaboration (M. H. Ahn {\it et al}.), Phys. Rev. Lett. {\bf 90}, (2003) 041801.
%
\bibitem{darkmatter}
K. G. Begeman, A. H. Broeils, and R. H. Sanders, Mon. Not. R. Astron,
Anatoly Klypin, John Holtzman, Joel Primack, Eniko Regos, Astrophys. J. {\bf 416}, (1993) 1; George R. Blumenthal, S.M. Faber, Joel R. Primack, Martin J. Rees, Nature {\bf 311}, (1984) 517; Robert K. Schaefer, Qaisar Shafi, Floyd W. Stecker, Astrophys. J. {\bf 347}, (1989) 575; Michael S. Turner, Published in *Asilomar 1998, Particle physics and the early universe* 113-128,  e-Print: astro-ph/9904051v1;
WMAP+BAO+SN, Recommended Parameter Values availuable in the address: \url{http://lambda.gsfc.nasa.gov/ pro-duct/map/dr3/parameters.cfm};
P. J. E. Peebles, Principles of Physical Cosmology, (Princeton University,
1993); Debasish Majumdar, e-Print: hep-ph/0703310;
A. Liddle, Introduction to Modern Cosmology, (John Wiley and Sons,
2003, 2.ed).
%
\bibitem{susyreviews}
H. E. Haber and G. L. Kane, Phys. Rept. {\bf 117}, (1985) 75; 
W. de Boer, Grand unified theories and supersymmetry in particle physics and cosmology, Prog. Part. Nucl. Phys. {\bf 33}, (1994) 201; 
S. P. Martin, A Supersymmetry primer, arXiv:hep-ph/9709356;  G. Kane, Perspectives on supersymmetry II, World Scientific (2010);
D. Kazakov, Supersymmetry on the Run: LHC and Dark Matter, Nucl. Phys. Proc. Suppl. {\bf 203-204}, (2010) 118.
%
\bibitem{susypapers}
P. Draper, P. Meade, M. Reece, D. Shih, Phys. Rev.  D{\bf 85},  (2012) 095007;
F. Mahmoudi, A. Arbey, M. Battaglia , A. Djouadi,  CERN-Conference, e-Print: arXiv:1211.2794;
M. S. Carena and H. E. Haber, Prog. Part. Nucl. Phys. {\bf 50},  (2003) 63;
Djouadi, A. et al. arXiv:1307.5205 [hep-ph];
Carena, M. et al. arXiv:1302.7033 [hep-ph].
\bibitem{ffv}
P. H.  Frampton,  Phys. Rev. Lett. {\bf 69}, (1992) 2889;
F. Pisano and V. Pleitez, Phys. Rev. D{\bf 46}, (1992) 410.
\bibitem{reduced}
J. G.  Ferreira, Jr, P. R. D.  Pinheiro, C. A. de  S.  Pires, P. S. Rodrigues da Silva, Phys. Rev. D{\bf84}, (2011) 095019. 
\bibitem{pheno}
V. T. N. Huyen, T. T.  Lam, H. N. Long, V. Q. Phong, e-Print: arXiv:1210.5833. 
 
\bibitem{DuongMa} T. V. Duong and E. Ma, Phys. Lett. B{\bf 316}, (1993) 307.

\bibitem{marcao} J. C. Montero, V. Pleitez, M. C. Rodriguez, Phys. Rev. D{\bf 65}, (2002) 035006.
 
\bibitem{susylong}
A similar SUSY version of this model was previously published in Ref.  D. T. Huong, L. T. Hue, M. C. Rodriguez, H. N. Long,  Nucl. Phys. B{\bf 870}, (2013) 293. 

\bibitem{mauro} D. Fregolente, M. D. Tonasse, Phys. Lett. B{\bf 555}, (2003) 7.

\bibitem{landaupole}
This is so because the model presents a Landau-pole at $4-5$~TeV energy scale, see Ref. A. G. Dias, R. Martinez, V. Pleitez,  Eur. Phys. J. C{\bf 39} (2005), 101; See also,  A. G. Dias, V. Pleitez, Phys. Rev. D{\bf 80}, (2009) 056007.

\bibitem{2fotons} 	
A. Alves, E. Ramirez Barreto, A. G. Dias, C. A. de S.Pires, F. S. Queiroz, P. S. Rodrigues da Silva, Phys. Rev. D{\bf 84}, (2011) 115004; A. Alves, E. Ramirez Barreto, A. G. Dias, C. A. de S. Pires, F. S. Queiroz, P. S. Rodrigues da Silva, Eur. Phys. J. C{\bf 73}, (2013) 2288; 
W. Caetano, C. A. de S. Pires, P. S. Rodrigues da Silva, D. Cogollo, Farinaldo S. Queiroz, 
e-Print: arXiv:1305.7246; Chong-Xing Yue, Qiu-Yang Shi, Tian Hua,
e-Print: arXiv:1307.5572.

\bibitem{potencialefetivosusy}
Y. Okada, M. Yamaguchi, T. Yanagida, Prog. Theor. Phys., 85 (1991), 1; H.E. Haber, R. Hempfling, Phys. Rev. Lett., {\bf 66}, (1991) 1815; J. Ellis, G. Ridolfi, F. Zwirner, Phys. Lett. B, B257 (1991), 83; J. Ellis, G. Ridolfi, F. Zwirner, Phys. Lett. B{\bf 262}, (1991) 477; A. Brignole, J. Ellis, G. Ridolfi, F. Zwirner, Phys. Lett., B{\bf 271}, (1991) 123.
	
\bibitem{finetuning} Edward Hardy, e-Print: arXiv:1306.1534 and references therein; Baer, Howard et al. arXiv:1306.2926 [hep-ph].

\end{thebibliography}
\end{document}